\documentclass[amsmath,amssymb,amsfonts,aps,pra,twocolumn,showpacs]{revtex4-1}
\usepackage{times}
\usepackage[pdftex]{graphicx}
\usepackage[utf8]{inputenc}
\usepackage{multirow}
\usepackage{color}
\usepackage[urlcolor=blue, hyperindex, colorlinks, bookmarks=true, linkcolor=blue, citecolor=blue]{hyperref}

\begin{document}

\title{Equivalence between an optomechanical system and a Kerr medium}

\author{Samuel~Aldana}
\author{Christoph~Bruder}
\author{Andreas~Nunnenkamp}
\affiliation{Department of Physics, University of Basel,
  Klingelbergstrasse 82, CH-4056 Basel, Switzerland}

\date{\today}

\begin{abstract}
We study the optical bistability of an optomechanical system in
which the position of a mechanical oscillator modulates the
cavity frequency. The steady-state mean-field equation of the
optical mode is identical to the one for a Kerr medium, and thus
we expect it to have the same characteristic behavior with a
lower, a middle, and an upper branch. However, the presence of
position fluctuations of the mechanical resonator leads to a new
feature: the upper branch will become unstable at sufficiently
strong driving in certain parameter regimes. We identify the
appropriate parameter regime for the upper branch to be stable,
and we confirm, by numerical investigation of the quantum
steady state, that the mechanical mode indeed acts as a Kerr
nonlinearity for the optical mode in the low-temperature limit.
This equivalence of the optomechanical system and the Kerr medium
will be important for future applications of cavity optomechanics
in quantum nonlinear optics and quantum information science.
\end{abstract}



\maketitle

\section{Introduction}

Photons are ideal carriers of quantum information \cite{OBrien2009}.
They can propagate large distances in optical fibers before being
absorbed, and their polarization has been used for quantum
communication and quantum information applications. However, 
photons barely interact, and thus it is difficult to
implement the quantum two-qubit gates needed for universal quantum
computation \cite{Kok2007RMP79}. This situation changes in an optical
medium where the photons can inherit an effective interaction, often
modeled as a Kerr nonlinearity.
This is why so-called Kerr media are important for quantum technology
based on photons \cite{Milburn1989PRL62, Chuang1996PRL76, Chuang1995PRA52, Braunstein2005RMP77}.

Recently, it was suggested that optomechanical systems
\cite{Aspelmeyer2013arx} operated in the single-photon strong-coupling
regime \cite{Nunnenkamp2011PRL107, Rabl2011PRL107} offer strong
effective photon-photon interactions. In optomechanical systems the
position of a mechanical oscillator modulates the properties and (most
commonly) the frequency of the optical cavity mode.  The radiation
pressure interaction is intrinsically nonlinear. It induces many
interesting effects and enables many applications, e.g.~sideband cooling
\cite{WilsonRae2007PRL99,Marquardt2007PRL99}, radiation-pressure shot
noise \cite{Caves1981,Braginsky1995,Borkje2010PRA82,Purdy2013Sci339},
photon blockade \cite{Rabl2011PRL107}, non-Poissonian photon statistics
and multiphoton transitions \cite{Kronwald2013PRA87},
and non-Gaussian and nonclassical mechanical states
\cite{Mancini1997PRA55, Bose1997PRA56, Nunnenkamp2011PRL107, Qian2012PRL109}.

\begin{figure}
\centering
\includegraphics[width=0.74\linewidth]{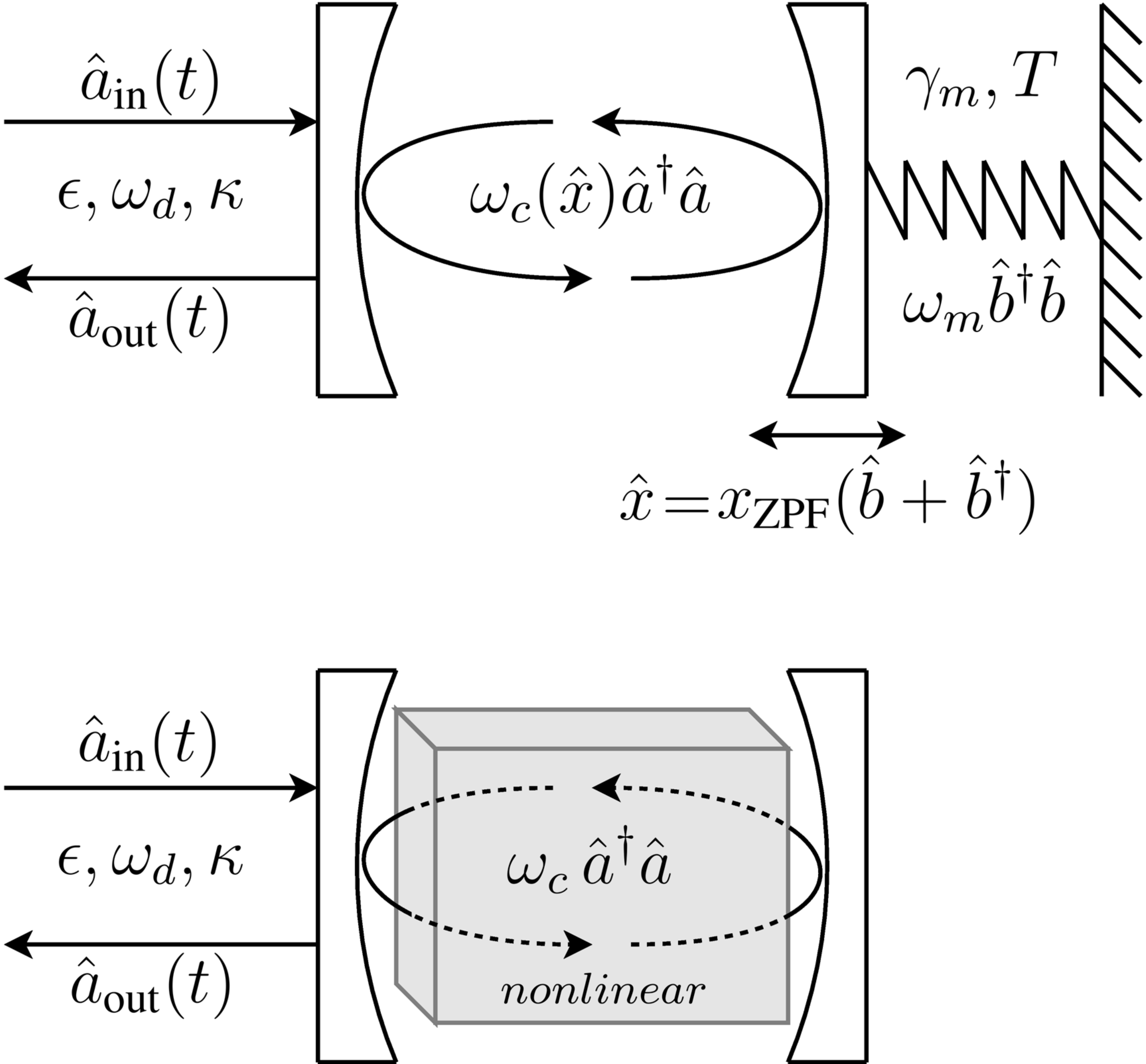}
\caption{Optomechanical setup (upper panel) and Kerr medium in a
  cavity (lower panel). The main part of the paper investigates in
  detail whether and in which way the two systems are equivalent.}
\label{fig:fig0}
\end{figure}

In this paper, we will focus on the phenomenon of optical bistability,
produced by the radiation pressure, and neglect
other nonlinear effects such as the photothermal effect
\cite{Braginsky1989PLA,Fomin2005JOS2,Marino2011PRE83,*Marino2013PRE87}
or a mechanical Duffing nonlinearity.
Under certain conditions and sufficiently strong driving there
are two classically stable equilibrium positions for the mechanical
oscillator and correspondingly for the optical cavity.
Optical bistability in optomechanical systems has been discussed in
the context of ponderomotive squeezing
\cite{Fabre1994PRA49, *Mancini1994PRA49} and entanglement
\cite{Ghobadi2011PRA84}, and led to one of the first experimental
observations of optomechanical coupling
\cite{Dorsel1983PRL51, Gozzini1985JOS}. Optical bistability has also
been discussed widely in the context of a Kerr medium
\cite{DrummondWalls1980JPA13,Walls2008}. 
This raises the question whether and in which way the optomechanical
system and the Kerr medium in a cavity can be considered to be
equivalent, see Fig.~\ref{fig:fig0} that shows both of these systems
schematically.  In the following we will investigate in detail the
similarities and differences between optical bistability in an
optomechanical system and a Kerr medium.

The paper is organized as follows.  In Sec.~\ref{sec:model} we
introduce the standard model of optomechanics -- a cavity whose
frequency is modulated by the position of a mechanical oscillator. We
briefly introduce the steady-state mean-field equations of the system
and the quantum Langevin description of quantum and thermal
fluctuations for a linearized radiation-pressure interaction. In
Sec.~\ref{sec:bistability} we show that the mean-field equation for
the optical mode is identical to the one for a Kerr medium, with a
lower, a middle and an upper branch. In the optomechanical system,
fluctuations of the mechanical mode change the picture. A study of the
stability of the different mean-field solutions against fluctuations
reveals a feature that is absent from the Kerr medium: the upper
branch becomes unstable for certain parameters. We derive conditions
on the parameters for this upper branch to remain stable.  The
stability requires the system to be in the resolved sideband regime
with a mechanical quality factor that is not too large.  In this case
we expect the mechanical resonator to act as an effective Kerr medium
for the optical mode, even in the quantum regime. This is confirmed in
Sec.~\ref{sec:equivalence}, where we compare the quantum steady states
of both the optomechanical system and the Kerr medium, obtained from
numerical solutions of the quantum master equations in the
low-temperature limit. The optomechanical system exhibits the expected
characteristic quantum signatures proving that it can be regarded as
an effective Kerr medium.

\section{Models for the optomechanical system and the Kerr medium}
\label{sec:model}

We first consider the standard model of optomechanics where the resonance
frequency of an optical cavity is modulated by the position of a
mechanical resonator (dispersive coupling). A monochromatic coherent
light field with frequency $\omega_d$ and amplitude $\epsilon$ drives
the optical mode. The full Hamiltonian, accounting for driving and
dissipation, is $\hat{H} = \hat{H}_0 + \hat{H}_d +\hat{H}_{\kappa} +
\hat{H}_{\gamma_m}$, where, in the rotating frame of the driving
($\hbar=1$),
\begin{equation}
\hat{H}_0 = \omega_m \hat{b}^{\dagger}\hat{b} - 
\Delta_0\hat{a}^{\dagger}\hat{a}  
- g_0 \hat{a}^{\dagger}\hat{a} (\hat{b}+\hat{b}^\dagger)\:,
\label{eq:Hamiltonian}
\end{equation}
and $\hat{H}_d = i \epsilon (\hat{a} - \hat{a}^{\dagger})$. Here,
$\hat{a}$ and $\hat{b}$ are the bosonic operators for the optical and
mechanical modes, $\Delta_0 = \omega_d-\omega_c$ is the detuning of the
drive from the unperturbed cavity resonance frequency $\omega_c$, and
$\omega_m$ the resonance frequency of the mechanical mode. The
optomechanical coupling is given by $g_0 = - x_{\text{ZPF}}
(\partial\omega_c/\partial x)$, where $x_{\text{ZPF}} =
(2M\omega_m)^{-1/2}$ is the zero-point fluctuation amplitude of the
mechanical resonator, $M$ its mass, and
$(\partial\omega_c/\partial x)$ is the derivative of the cavity
frequency with respect to the resonator position
$\hat{x}=x_{\text{ZPF}}(\hat{b}+\hat{b}^{\dagger})$. 
The term $\hat{H}_{\kappa}$ describes the damping of the optical 
cavity at rate $\kappa$, and $\hat{H}_{\gamma_m}$ 
the damping of the mechanical resonator at rate $\gamma_m$.
This leads to the definition of two important ratios,
the sideband parameter $\omega_m/\kappa$ and the
mechanical quality factor $Q_m=\omega_m/\gamma_m$.

Using the input-output formalism \cite{Walls2008, Clerk2010RMP82},
the dissipative dynamics of the system is described by the quantum
Langevin equations (QLEs)
\begin{subequations}
\label{eq:QLE}
\begin{eqnarray}
\dot{\hat{a}} &=& \left(i \Delta_0 - \frac{\kappa}{2} \right) \hat{a} 
+ i g_0 \hat{a}
(\hat{b} + \hat{b}^\dagger) -\sqrt{\kappa}\, \hat{a}_{\text{in}}\:,
\label{eq:QLEa}
\\
\dot{\hat{b}} &=& - \left(i \omega_m + \frac{\gamma_m}{2} \right)
\hat{b} 
+ i g_0 \hat{a}^\dagger \hat{a} -\sqrt{\gamma_m}\, \hat{\eta}\:,
\label{eq:QLEb}
\end{eqnarray}
\end{subequations}
where $\hat{a}_{\text{in}}(t) = \bar{a}_{\text{in}} + \hat{\xi}(t)$
consists of a coherent driving amplitude $\bar{a}_{\text{in}} =
\epsilon/\sqrt{\kappa}$ and a vacuum noise operator $\hat{\xi}$ which
satisfies $\langle \hat{\xi}(t)\hat{\xi}^{\dagger}(t') \rangle =
\delta(t-t')$ and $\langle \hat{\xi}^{\dagger}(t)\hat{\xi}(t')\rangle=
0$. Similarly, the noise operator $\hat{\eta}$ describes
coupling to a Markovian bath at temperature $T$,
i.e., $\langle
\hat{\eta}(t)\hat{\eta}^{\dagger}(t') \rangle =
(n_{\text{th}}+1)\delta(t-t')$ and 
$\langle \hat{\eta}^{\dagger}(t)\hat{\eta}(t') \rangle =
n_{\text{th}}\delta(t-t')$. 
In the absence of any other coupling, the bath gives rise to a thermal state
with mean occupation number $n_{\text{th}} = [\exp(\omega_m/k_B T)
-1]^{-1}$ for the mechanical oscillator.
This treatment of the mechanical dissipation in the form of a QLE for
the mechanical amplitude $\hat{b}$, rather than for the displacement $\hat{x}$,
is correct as long as $Q_m\gg 1$.

The optical and mechanical field operators can be split into
a coherent mean-field amplitude and fluctuations:
$\hat{a}(t) = \bar{a} + \hat{d}(t)$ and $\hat{b}(t) = \bar{b} +
\hat{c}(t)$. Inserting these expressions in the QLEs~\eqref{eq:QLE}, we
obtain two coupled mean-field equations (MFEs) for the 
amplitudes $\bar{a}$ and $\bar{b}$. In steady state they read
\begin{subequations}
\label{eq:MFE}
\begin{eqnarray}
0 &=& \left[i\Delta_0 + ig_0 \left(\bar{b}+{\bar{b}}^{*} \right) 
- \frac{\kappa}{2} \right] \bar{a} - \epsilon\:,
\label{eq:MFEa}
\\
0 &=& - \left( i\omega_m + \frac{\gamma_m}{2} \right) \bar{b} 
+ ig_0 |\bar{a}|^2\:.
\label{eq:MFEb}
\end{eqnarray}
\end{subequations}
The coherent amplitude of the optical field $\bar{a}$ corresponds to a
mean cavity occupation $\bar{n}=|\bar{a}|^2$ and produces a static
radiation-pressure force $g_0\,\bar{n}/x_{\text{ZPF}}$ on the
resonator, displacing its equilibrium position by an amount
$x_{\text{ZPF}}(\bar{b}+\bar{b}^{*})$. Proceeding this way we
eliminate the coherent drive $\epsilon$ from the QLEs for the
operators $\hat{c}$ and $\hat{d}$ which describe thermal and quantum
fluctuations around the mean-field values.

For large optical mean-field amplitudes $|\bar{a}|\gg 1$ and small
coupling $g_0\ll \kappa,\omega_m$, we can neglect the nonlinear terms
like $\hat{d}^{\dagger}\hat{d}$ or $\hat{d}\hat{c}$ in the QLEs.
As a result, the optomechanical interaction becomes bilinear:
$g_0\hat{a}^{\dagger}\hat{a}(\hat{b}+\hat{b}^{\dagger}) \to g_0
(\bar{a}^{*}\hat{d}+\bar{a}\,\hat{d}^{\dagger})(\hat{c}+\hat{c}^{\dagger})$. 
Introducing the convenient vector notation $\boldsymbol{\hat{u}} =
(\hat{d}^{\dagger},\hat{d},\hat{c}^{\dagger},\hat{c})^T$ and
$\boldsymbol{\hat{u}}_{\text{in}} =
(\sqrt{\kappa}\hat{\xi}^{\dagger},\sqrt{\kappa}\hat{\xi},
\sqrt{\gamma_m}\hat{\eta}^{\dagger},\sqrt{\gamma_m}\hat{\eta})^T$,
we can write the linearized QLEs in matrix form,
\begin{equation}
\frac{d}{dt} \boldsymbol{\hat{u}}(t) = -\boldsymbol{A} \cdot
\boldsymbol{\hat{u}}(t) - \boldsymbol{\hat{u}}_{\text{in}}(t)\:,
\label{eq:linQLE}
\end{equation}
where $\boldsymbol{A}$ reads
\begin{equation}
\boldsymbol{A} = 
\begin{pmatrix}
\frac{\kappa}{2}+i\Delta & 0 & ig^{*} & ig^{*} \\
0 & \frac{\kappa}{2}-i\Delta & -ig & -ig \\
ig & ig^{*} & \frac{\gamma_m}{2}-i\omega_m & 0 \\
-ig & -ig^{*} & 0  & \frac{\gamma_m}{2}+i\omega_m
\end{pmatrix}.
\label{eq:matrixA}
\end{equation}
The new parameters entering the matrix $\boldsymbol{A}$ are the
enhanced optomechanical coupling $g = g_0 \bar{a}$ and the effective
detuning $\Delta = \Delta_0 + g_0(\bar{b}+\bar{b}^{*}) = \Delta_0 +
2\bar{n}g_0^2/\omega_m$.

The Kerr medium \cite{DrummondWalls1980JPA13,Walls2008}, to which we aim
to compare the optomechanical system, is described by the Hamiltonian
$\hat{H}' = \hat{H}_K + \hat{H}_d + \hat{H}_{\kappa}$, where, in the
rotating frame of the driving,
\begin{subequations}
\label{eq:HamiltonianKerr}
\begin{eqnarray}
\hat{H}_K &=& -\Delta_0 \hat{a}^{\dagger}\hat{a} - \frac{g_0^2}{\omega_m}
\left(\hat{a}^{\dagger}\hat{a}\right)^2\:, \label{eq:HamiltonianKerra}
\\
\hat{H}_d &=& i\epsilon(\hat{a}-\hat{a}^{\dagger})\:, 
\label{eq:HamiltonianKerrb}
\end{eqnarray}
\end{subequations}
and $\hat{H}_{\kappa}$ describes again the damping of the optical
cavity at rate $\kappa$.  The QLE for this optical mode $\hat{a}$ is
\begin{equation}
\dot{\hat{a}} = \left[ i \left( \Delta_0+\frac{g_0^2}{\omega_m}\right) 
- \frac{\kappa}{2} \right]\hat{a} + 
2 i \frac{g_0^2}{\omega_m} \hat{a}^{\dagger} \hat{a}^2 
- \sqrt{\kappa}\hat{a}_{\text{in}}\:, 
\label{eq:QLEKerr}
\end{equation}
where the input operator $\hat{a}_{\text{in}}(t)$ is the same as
for the optomechanical system. The steady-state equation for the
mean-field amplitude $\bar{a}$ is
\begin{equation}
0 = \left[ i \left( \Delta_0 + \frac{g_0^2}{\omega_m} \right) 
- \frac{\kappa}{2} \right] \bar{a} + 
2 i \frac{g_0^2}{\omega_m} |\bar{a}|^2 \bar{a} - \epsilon\:. 
\label{eq:MFEKerr}
\end{equation}

Replacing $\Delta_0$ by $\Delta_0-g_0^2/\omega_m$ in
Eq.~\eqref{eq:MFEKerr} yields the equation for the optical mean-field
amplitude $\bar{a}$ of the optomechanical system obtained from
Eq.~\eqref{eq:MFE} by eliminating the mechanical mean-field amplitude
$\bar{b}$. This frequency shift of the detuning $\Delta_0$ is
consistent with the fact that $\hat{H}_0$ and $\hat{H}_K$ are
connected by the canonical (polaron) transformation $\hat{U} =
\exp[(g_0/\omega_m) (\hat{b}-\hat{b}^{\dagger})
\hat{a}^{\dagger}\hat{a}]$. Applying $\hat{U}$ to the optomechanical
Hamiltonian $\hat{H}_0$, Eq.~\eqref{eq:Hamiltonian}, we obtain
$\hat{U}\hat{H}_0\hat{U}^{\dagger} = \hat{H}_K + \omega_n
\hat{b}^{\dagger}\hat{b}$. In this frame, the optomechanical
interaction is eliminated and the optical mode acquires a Kerr
nonlinearity of the form of Eq.~\eqref{eq:HamiltonianKerra}
\cite{Nunnenkamp2011PRL107, Rabl2011PRL107}.

\section{Optical bistability in the semiclassical regime}
\label{sec:bistability}

In the following, we will first show that the optomechanical system
has MFEs with three solutions in a certain range of
driving frequency and driving amplitude, just as the Kerr medium
does. After discussing the characteristic behavior of the mean-field
solutions in the regime of optical bistability, we study the
stability of the mean-field solutions against fluctuations of both
the optical and mechanical mode and point out the differences with the
Kerr medium. Finally, we find parameters for which the optomechanical
system is accurately described by an effective Kerr medium.

\subsection{Bistability at the mean-field level}

\begin{figure*}
\centering
\includegraphics[width=0.9\linewidth]{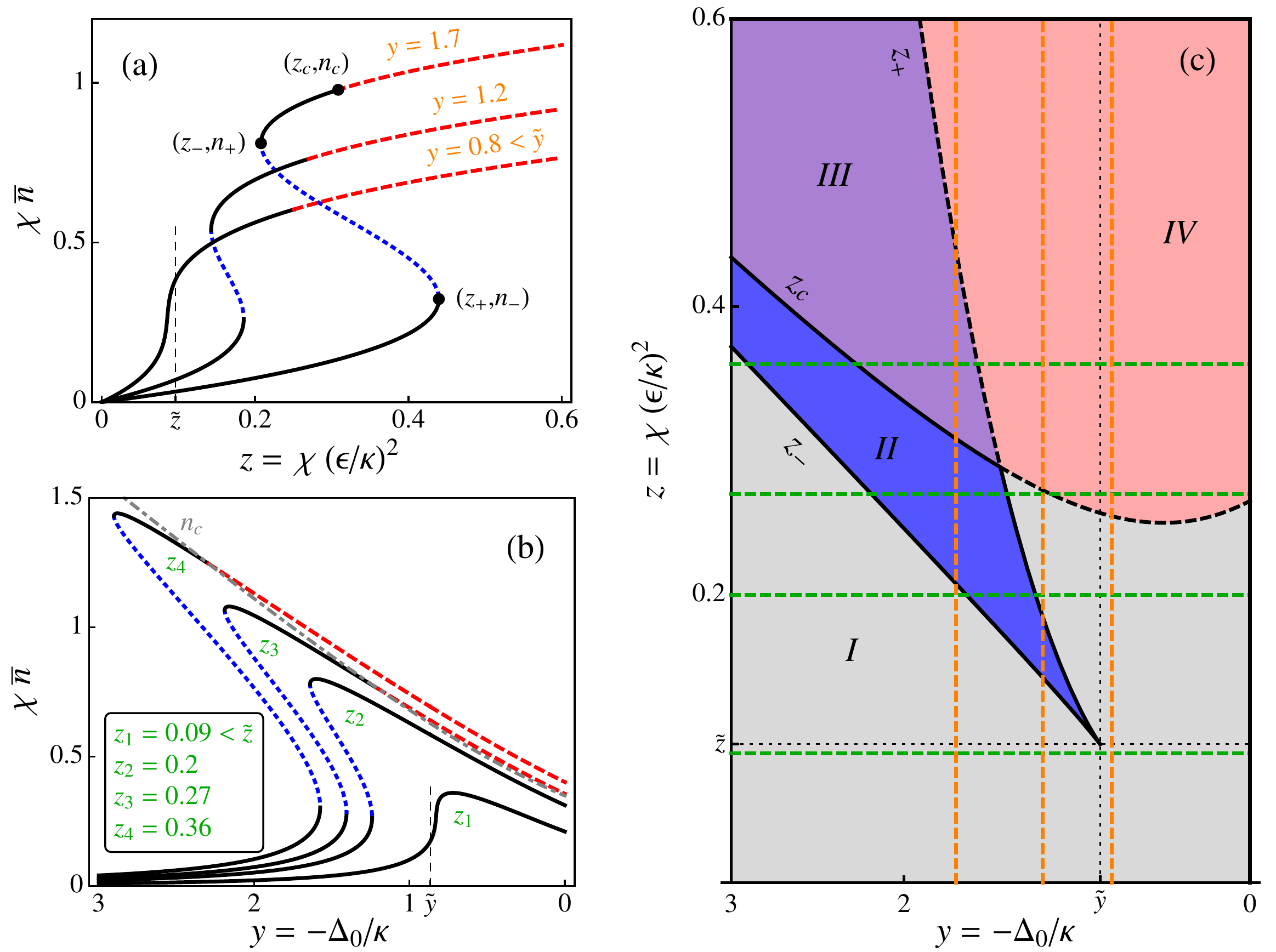}
\caption{(Color online) \textit{Optical bistability in the
  semiclassical regime}. Typical curves for the mean-field cavity
  occupation $\bar{n}$ as
  a function of the dimensionless driving power $z$ (a) and the
  dimensionless detuning $y$ (b), obtained from the condition $p(\chi\bar{n})=0$
  [see Eq.~\eqref{eq:rootequation}]. According to the stability criteria
  $c_{1,2}>0$ [see Eqs.~\eqref{eq:RHcriteria}], Gaussian fluctuations
  lead to stable (solid black) or unstable (dotted blue and dashed
  red) mean-field solutions. As in the case of the Kerr medium, the first
  criterion $c_1>0$ always yields an unstable middle branch (dotted
  blue), while the additional criterion for the optomechanical system
  $c_2>0$ can turn part of the upper or only branch unstable (dashed
  red). In (b) we also show the critical mean-field occupation
  $n_c$ (dash-dotted gray) obtained from the condition $c_2 = 0$.
  In (c) we summarize the behavior of the mean-field
  solution as a function of the parameters $y$ and $z$.
  In regions $II$ and $III$, between the curves $z_-$ and $z_+$,
  Eqs.~\eqref{eq:threshold} and \eqref{eq:zcondition} are satisfied
  and there are three distinct mean-field solutions; the middle branch
  is always unstable. In region $II$ (blue) the lower and upper branches
  are stable. In region $III$ (purple) the second stability
  criterion shows the upper branch to be unstable ($c_2<0$) and only
  the lower branch is stable. In regions $I$ and $IV$ the mean-field
  equations (MFEs) have only one solution. Below the $z_c$ curve in region
  $I$ (gray) this unique branch is stable, while in region $IV$ (red)
  the second criterion again shows that this solution is unstable
  ($c_2<0$). The values of the detuning $y$ and driving power $z$
  used in (a) and (b) are indicated by the orange and green dashed lines.
  Note that none of these features depends on the nonlinearity
  parameter $\chi$, due to appropriate scaling of the axes. The threshold
  detuning $\tilde{y}$ and driving power $\tilde{z}$ indicate the minimal
  values of $y$ and $z$ needed for the MFEs to have three
  solutions. The sideband parameter and mechanical quality factor
  chosen to show the influence of the second stability criterion
  $c_2>0$ are $\omega_m/\kappa=10$ and $Q_m=1000$.}
\label{fig:fig1}
\end{figure*}

We briefly review the origin of bistability in the mean-field
equations of the optomechanical system 
\cite{Meystre1985JOS2,Meystre2007,Gozzini1985JOS,Fabre1994PRA49,*Mancini1994PRA49}.

To simplify the notation we define the dimensionless nonlinearity
parameter $\chi$, detuning $y$, and driving power $z$ by
\begin{align*}
\chi &= \frac{g_0^2}{\omega_m\kappa}\:,\\
y &= -\frac{\Delta_0}{\kappa}\:,\\
z &= \chi \left(\frac{\epsilon}{\kappa}\right)^2\:.
\end{align*}
Combining Eqs.~\eqref{eq:MFEa} and \eqref{eq:MFEb} we obtain a
third-order polynomial root equation for the mean-field cavity
occupation, $p(\chi\bar{n})=0$, where
\begin{equation}
p(\lambda) = 4\lambda^3 - 4 y \lambda^2 
+ \left( y^2 + \frac{1}{4} \right) \lambda - z\:.
\label{eq:rootequation}
\end{equation}
The MFE for the Kerr medium, Eq.~\eqref{eq:MFEKerr},
leads to the same equation for $\bar{n}$, provided we
replace $y$ by $y-\chi$ in Eq.~\eqref{eq:rootequation}.

Equation~\eqref{eq:rootequation} indicates that the MFEs can have
either one or three solutions, depending on the number of real roots
of the polynomial. The three roots depend on the dimensionless
detuning $y$ and driving power $z$. Since the mean-field cavity
occupation $\bar{n}$ follows from $p(\chi\bar{n})=0$, the nonlinearity
parameter $\chi$ determines whether optical bistability occurs at
small or large driving power and photon number.

The optical mean-field amplitude is $\bar{a} = - e^{i\varphi}
\sqrt{\lambda/\chi}$, where $\varphi = \arctan(4\lambda-2y)$. If the
detuning $y$ and driving power $z$ are such that the equation
$p(\lambda)=0$ has three real roots, the smaller $\chi$, the more
distant in phase space are the different optical mean-field amplitudes
$\bar{a}$. A similar observation can be made concerning the mechanical
resonator: the equation $p(\lambda) = 0$ also holds for $\lambda =
\sqrt{\chi \omega_m/(4\kappa)}(\bar{b}+\bar{b}^{*})$, where
$\bar{b}+\bar{b}^{*}$ is the equilibrium position of the mechanical
resonator in units of $x_{\text{ZPF}}$. Therefore, the smaller $\chi$
and the sideband parameter $\omega_m/\kappa$, the more distant are the
different equilibrium positions.

We now examine some characteristic features of the MFEs, which occur
both in an optomechanical system (\ref{eq:MFE}) and a Kerr medium
(\ref{eq:MFEKerr}). To this end, we find the conditions on the
detuning $y$ and the driving power $z$ for the MFEs to have three
solutions, and illustrate them with a few examples.

First we observe that the equation $p(\lambda)=0$ can have three real roots
only if the detuning $y$ and the driving power $z$ exceed some
threshold value $\tilde{y}$ and $\tilde{z}$
 \cite{Risken1987PRA35,*Vogel1989PRA39,Fabre1994PRA49,*Mancini1994PRA49},
\begin{subequations}
\label{eq:threshold}
\begin{eqnarray}
y > \tilde{y} &=& \frac{\sqrt{3}}{2} \simeq 0.87\:, \\
z > \tilde{z} &=& \frac{1}{6\sqrt{3}} \simeq 0.1\:.
\end{eqnarray}
\end{subequations}
Therefore, optical bistability can only be found for red-detuned
driving frequencies. In addition, the three roots are real only if
\begin{equation}
z_-(y) < z < z_+(y)\:,
\label{eq:zcondition}
\end{equation}
where
\begin{equation*}
z_{\pm}(y) = \frac{1}{27}\left[ y (y^2 + 3\tilde{y}^2) \pm (y^2 -
\tilde{y}^2 )^{3/2}\right]\:.
\end{equation*}

The region in $(y,z)$-parameter space where Eqs.~\eqref{eq:threshold}
and \eqref{eq:zcondition} are satisfied is shown in
Fig.~\ref{fig:fig1}(c) with the labels \textit{II} (blue) and
\textit{III} (purple).
In this region the three mean-field occupations
satisfy $\bar{n}_1 < n_- < \bar{n}_2 < n_+ < \bar{n}_3$,
where $n_{\pm}$ are found from $p'(\chi n_{\pm}) = 0$ and read
\begin{equation}
\chi n_{\pm}(y) = \frac{1}{6}\left[2y \pm (y^2-\tilde{y}^2)^{1/2}\right].
\label{eq:turningpoints}
\end{equation}
In the following, we refer to $\bar{n}_1$, $\bar{n}_2$, and
$\bar{n}_3$ as the lower, middle, and upper branch of the MFEs.

In Fig.~\ref{fig:fig1}(a) we show the mean-field
occupation $\chi\bar{n}$ as a function of the driving
power $z$ for fixed detuning $y$.
For an increasing driving power $z$ and a detuning above
the threshold $y>\tilde{y}$, the three branches of the mean-field
occupation $\bar{n}$ form a characteristic $S$-shaped curve. The lower
branch starts from the origin and ends at the turning point given by
$(z_+,n_-)$ where the middle branch starts. The upper branch starts
from the second turning $(z_-,n_+)$, where the middle branch ends,
and increases further.

In Fig.~\ref{fig:fig1}(b) we plot the mean-field occupation
$\chi\bar{n}$ as a function of the detuning $y$ for fixed driving
power $z$.  The cavity line shape is approximately Lorentzian if the
driving power is far below the threshold $z\ll\tilde{z}$ (not shown).
For larger and larger $z$ it becomes more and more asymmetric and tilts until
for $z=\tilde{z}$, it has an infinite slope at
$y=\tilde{y}$. For a driving power beyond this threshold
the cavity line-shape has three branches in the range
of detuning $y$ determined by Eq.~\eqref{eq:zcondition}.

According to these considerations, the optomechanical system and the
Kerr medium are equivalent at the level of the steady-state MFEs.  Our
next goal is to discuss the stability of the different branches of the
MFEs.  The existence of three solutions to the MFEs indicates that the
optomechanical system may be in a regime of bistability, with stable
lower and upper branches, as well as an unstable middle branch. While
for the Kerr medium this is always true \cite{DrummondWalls1980JPA13},
a stability analysis leads to different conclusions in the case of the
optomechanical system. In addition, if the detuning $y$ and driving
power $z$ lead to a unique solution for the mean-field cavity
occupation $\bar{n}$, this solution is always stable for the Kerr
medium, but not necessarily so for the optomechanical system.

\subsection{Stability analysis of the mean-field solutions}

The upper and lower branches are always stable for the Kerr medium.
To find the range of parameters where the optomechanical system
reproduces this behavior, we analyze the stability
of the different branches of the MFEs~\eqref{eq:MFE} against
fluctuations of both the optical and mechanical modes.

The stability of a point in any of the branches of the MFEs is
established, if the linear QLEs~\eqref{eq:linQLE}, describing the
fluctuations around this point, are stable. 
This in turn is ensured if all the eigenvalues of the matrix $\boldsymbol{A}$ given in
Eq.~\eqref{eq:matrixA}, derived from the corresponding mean-field
amplitudes $\bar{a}$ and $\bar{b}$, have positive real parts. 
This has to be verified even if the MFEs have only one solution.

The differences and similarities between the optomechanical system and
the Kerr medium are summarized in Table~\ref{tab:diff}.

\begin{table}[!h]
\caption{Stability for the different branches in an optomechanical
system and a Kerr medium determined from the QLEs~\eqref{eq:linQLE}
and \eqref{eq:QLEKerr}. The critical mean-field occupation $n_c$ is
found from the stability criterion, Eq.~\eqref{eq:RH2},
and depends on the detuning $y=-\Delta_0/\kappa$, the sideband parameter
$\omega_m/\kappa$, and the mechanical quality factor $Q_m$.}
\begin{center}
\begin{tabular}{ccccc}
\hline
\multicolumn{2}{c}{Branch} & \multirow{2}{*}{\: Kerr medium \:} 
& \multicolumn{2}{c}{\multirow{2}{*}{Optomechanical system}}\\
\cline{1-2}
\: \# \: & type &  & \\
\hline
\multirow{4}{*}{3} & lower & stable & \multicolumn{2}{c}{stable} \\ 
& \: middle \: & unstable & \multicolumn{2}{c}{unstable} \\
& \multirow{2}{*}{upper} & \multirow{2}{*}{stable} 
& \quad stable\quad & \quad unstable \\[-0.2em]
& & & \quad$\bar{n}<n_c$\quad & \quad$\bar{n}>n_c$\quad \\
\hline
\multirow{2}{*}{1} & \multirow{2}{*}{-} & \multirow{2}{*}{stable} 
& \quad stable\quad & \quad unstable \\[-0.2em]
& & & \quad$\bar{n}<n_c$\quad & \quad$\bar{n}>n_c$\quad \\
\hline
\end{tabular}
\end{center}
\label{tab:diff}
\end{table}

The difference between the two systems is explained by the parametric
instability in the optomechanical system
\cite{Marquardt2006PRL96, Ludwig2008NJP10} that occurs
at a mean-field occupation $\bar{n}$ above some critical value $n_c$.
Around such a mean-field solution, the linear dynamics of
optical and mechanical fluctuations becomes unstable.
This particular feature of the optomechanical system is
illustrated in Fig.~\ref{fig:fig1}; it is absent for the Kerr medium.

In Figs.~\ref{fig:fig1}(a) and \ref{fig:fig1}(b), we indicate the
unstable segments of the branches where $\bar{n}>n_c$. 
In case the MFEs have three branches, this critical value for the
mean-field occupation $n_c$ systematically lies in the upper branch or
in its extension to the region where there is only one branch.

In Fig.~\ref{fig:fig1}(a), for a fixed detuning above threshold
$y>\tilde{y}$, the upper branch is stable only in a finite segment
near the second turning point $n_+$ at the beginning of the upper
branch. The size of this stable segment diminishes as the detuning $y$
increases, and shrinks to a single point in the limit of a far
red-detuned driving frequency. The same effect is seen in
Fig.~\ref{fig:fig1}(b). With increasing driving power $z$ the
stability in the upper branch is confined to a smaller and smaller
segment near the maximum of the cavity line shape.

In Fig.~\ref{fig:fig1}(c), the regions in $(y,z)$-parameter space
where the upper or only branch turns unstable are labeled by \textit{III}
and \textit{IV}. These are the regions where the driving power $z$ is
larger than the critical value $z_c$, found by
solving the equation $p\left(\chi n_c\right)=0$ for $z$, where $p$ is given
in Eq.~\eqref{eq:rootequation}. The range of detuning $y$ or driving
power $z$ at which bistability is observed shrinks with increasing $y$
or $z$.

We now characterize the regime leading to optical bistability in the
optomechanical system, and therefore examine how the stability of
the branches depends on the parameters. To this end, we apply
the Routh-Hurwitz criterion to the linear QLEs~\eqref{eq:linQLE}. 
Two conditions have to be satisfied for a particular mean-field solution
to be stable, $c_{1,2} >0$, where 
\footnote{In 
Refs.~\onlinecite{Genes2008PRA78,Vitali2007PRL98,*Vitali2007PRA76,Genes2008PRA77},
similar criteria have been obtained using a quantum Brownian motion
approach to treat mechanical dissipation. Their criteria are
equivalent to $c_{1,2}$ in the limit $Q_m \gg 1$.}
\begin{subequations}
\label{eq:RHcriteria}
\begin{eqnarray}
c_1 &=& 4 |g|^2 \Delta + \omega_m \left( \Delta^2 + \frac{\kappa^2}{4}\right), 
\label{eq:RH1}
\\
c_2 &=& \kappa\; \gamma_m \left[ \left(\Delta^2 - \omega_m^2\right)^2 
+ \frac{1}{2} \left(\Delta^2 + \omega_m^2 \right) \left(\kappa 
+ \gamma_m \right)^2 \right. \nonumber
\\
& & \; \left. + \frac{1}{16} \left(\kappa + \gamma_m \right)^4 \right]
- 4 |g|^2 \Delta \,\omega_m \left(\kappa +\gamma_m\right)^2.
\label{eq:RH2}
\end{eqnarray}
\end{subequations}

The identification of the parameter regime leading to $c_{1,2} > 0$ is
done as follows.  We replace $|g|^2$ and $\Delta$ by their
$\bar{n}$-dependent expressions,
\begin{align*}
|g|^2 &= \kappa \omega_m\, \chi\bar{n}\,, \\
\Delta &= \kappa(2\chi\bar{n}-y)\:,
\end{align*}
in Eqs.~\eqref{eq:RHcriteria}, and express $c_{1,2}$ as functions of
the rescaled mean-field occupation $\chi\bar{n}$, the detuning $y$,
the sideband parameter $\omega_m/\kappa$, and the mechanical quality
factor $Q_m=\omega_m/\gamma_m$.

From the condition $c_1 < 0$ we conclude that the middle branch is unstable
\cite{Meystre1985JOS2, Meystre2007, Fabre1994PRA49, *Mancini1994PRA49}. 
This follows from 
$\text{sgn}(c_1) = \text{sgn}\left[(n_{+} -
  \bar{n})(n_{-}-\bar{n})\right]$, 
where $n_{\pm}$, Eq.~\eqref{eq:turningpoints}, are the values of the
mean-field cavity occupation at the lower and upper limits of the
middle branch.  The physical interpretation of this condition is
simple.  In the middle branch, the modification of the mechanical
frequency due to radiation pressure, also known as the optical spring
effect, is such that the modified mechanical force is no longer a
restoring force.

In the Kerr medium, the same stability condition, $c_1>0$, is found
from the linear QLEs, obtained by substituting
$\hat{a} = \bar{a} + \hat{d}$ in Eq.~\eqref{eq:QLEKerr}
and neglecting second- and third-order terms in $\hat{d}$,
$\hat{d}^{\dagger}$. No other criteria are needed to establish 
the stability of the system, and therefore the lower and upper
branches are always stable.

The condition $c_2 = 0$ is equivalent to the relaxation rate of the
system going to zero \cite{Genes2008PRA78}. In a stable system, this
relaxation rate is the real part of the eigenvalue of $\boldsymbol A$
closest to zero.  Above the critical mean-field occupation,
$\bar{n}>n_c$, this relaxation rate becomes negative, $c_2<0$, and the
branch turns unstable. If in addition $\bar{n}$ is the only mean-field
solution, the system is parametrically unstable.  We find $n_c$ by
solving the equation $c_2=0$ for $\bar{n}$, as a function of the
detuning $y$, the sideband parameter $\omega_m/\kappa$, and the
mechanical quality factor $Q_m$.

It turns out that $n_c$ always lies in the upper branch or in its
extension to the region with only one branch. This can be seen as
follows.  Since the condition $c_2 > 0$ is automatically satisfied for
negative \textit{effective} detuning, $\Delta \leq 0$, we find a lower
bound for the critical occupation,
\begin{equation*}
n_c \; \geq \; n_{\Delta} = \frac{y}{2\chi}\:.
\end{equation*}
In addition, the effective detuning $\Delta$ always turns positive in
the upper branch, since $n_{\Delta} \geq n_+$. Thus the upper branch
is only stable in the range $n_+ < \bar{n} < n_c$. This
stable portion can be very small, e.g., in the extreme case $-\Delta_0
\gg \kappa$ and $\gamma_m = 0$, we have $n_c = n_{\Delta} \simeq n_+$.

\begin{figure}
\centering
\includegraphics[width=\linewidth]{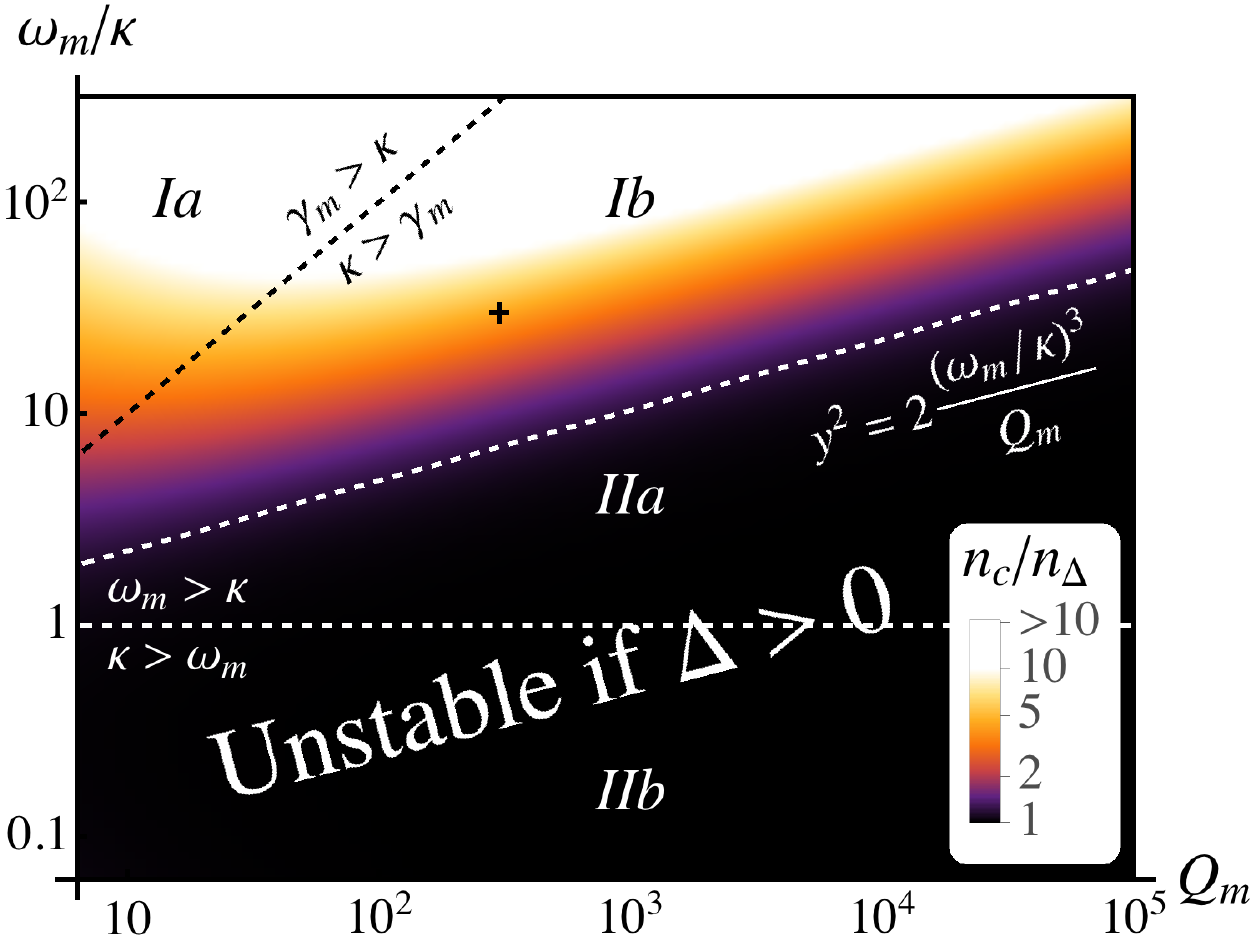}
\caption{(Color online) Critical cavity occupation $n_c$
  in units of $n_{\Delta}$, as a function of the sideband parameter
  $\omega_m/\kappa$ and the mechanical quality factor $Q_m$.  At
  $n_c$ the mean-field solution $\bar{n}$ leads to unstable linear
  dynamics for the optomechanical system. The cavity occupation
  $n_{\Delta} = y/(2\chi)$ marks the point at which the effective
  detuning $\Delta$ becomes positive. We find $n_c$ from
  the second stability criterion, Eq.~\eqref{eq:RH2}. The bare
  detuning is $y = -\Delta_0 / \kappa = 1.5$. Note that the
  ratio $n_c/n_{\Delta}$ does not depend on the nonlinearity parameter
  $\chi$. The black cross indicates the parameters used in
  Fig.~\ref{fig:fig3}.}
\label{fig:fig2}
\end{figure}

In Fig.~\ref{fig:fig2} we compare the critical mean-field cavity
occupation $n_c$ to the occupation $n_{\Delta}$ at which $\Delta$ changes sign.
The ratio $n_c/n_{\Delta}$ is shown as a function of $\omega_m/\kappa$
and $Q_m$.  If $n_c/n_{\Delta}$ is large, the upper branch is stable
beyond the parameter range leading to bistability, $n_c \gg n_+$,
mimicking the behavior of the Kerr medium. On the contrary, if
$n_c/n_{\Delta}\simeq 1$, the upper branch turns unstable for $\Delta
> 0$ and is only stable on a finite segment near its beginning.

We can distinguish four parameter regimes which encompass most
experimental situations.

\subsubsection{Resolved sideband and large mechanical damping (Ia)}

For extremely low cavity damping, $\omega_m > \gamma_m > \kappa$, the
critical occupation $n_c$ is approximately
\begin{equation*}
\chi n_c = \frac{1}{4}\left(y + \sqrt{y^2 + 
2\, Q_m \frac{\omega_m}{\kappa}}\right)\:.
\end{equation*}
In the case of a fixed detuning satisfying $y^2 \ll 2 Q_m
\omega_m/\kappa$, we have $n_c \gg n_{\Delta}$ and the upper
branch is stable on a considerable segment, extending up to driving
powers $z$ and mean-field occupations $\bar{n}$ that are much larger
than those needed for bistable MFEs, i.e., $z_c \gg z_+$ and
$n_c \gg n_+$. We recall that $z_c$ is found by solving the equation
$p(\chi n_c)=0$, with $p$ defined in
Eq.~\eqref{eq:rootequation}. Therefore, the mean-field behavior of the
optomechanical system is equivalent to the behavior of a Kerr medium
in the regime of bistability.
In Ref.~\onlinecite{Kronwald2013PRA87}, the optomechanical system was
compared to the Kerr medium in terms of the full counting statistics
of photons. Although the two systems can behave differently in some
regime of parameters, the authors demonstrate that the influence of
the mechanical resonator reduces to an effective Kerr nonlinearity
when $\gamma_m\sim\kappa$, in particular with $y=\omega_m/\kappa$.

\subsubsection{Resolved sideband and small mechanical damping (Ib and IIa)}

In the regime characterized by $\omega_m > \kappa > \gamma_m$, the
critical mean-field cavity occupation is found to be approximately
\begin{equation}
\chi n_c = \frac{1}{4}\left(y + \sqrt{y^2 + 
2\, \frac{(\omega_m/\kappa)^3}{Q_m}}\right).
\label{eq:criticalMF2}
\end{equation}
In this case, the parameter $(\omega_m/\kappa)^3/Q_m$ plays an
important role to characterize the mean-field behavior.

If $Q_m > (\omega_m/\kappa)^3$, we obtain $n_c \simeq n_{\Delta}$
for a detuning above the bistability threshold $y > \tilde{y}$. 
In this case, the upper branch turns unstable if the effective
detuning is positive, $\Delta >0$. In addition, this means that if the
detuning is negative and large, such that $-\Delta_0 \gg \kappa$, the
stable segment is small, as $n_{\Delta} \simeq n_+$.

In the opposite limit, $Q_m \ll (\omega_m/\kappa)^3$, we can have $n_c
\gg n_{\Delta}$ as in the previous case ($\gamma_m>\kappa$), provided
the detuning $y$ satisfies $y^2 \ll (\omega_m/\kappa)^3/Q_m$. The same
conclusions then apply, i.e., $z_c \gg z_+$ and $n_c \gg
n_+$, and the mean-field behavior of the optomechanical system and the
Kerr medium is equivalent in the parameter regime of bistability.

Using the exact expression for $n_c$, we see in Fig.~\ref{fig:fig2}
that the border between the region where the optomechanical system
experiences a parametric instability as soon as $\Delta>0$ (black
region), and the region where the system is still linearly stable for
some positive effective detuning, $n_c > n_{\Delta}$, is approximately
given by $y^2=2(\omega_m/\kappa)^3/Q_m$. Above this line, an
optomechanical system driven to the regime of bistability behaves like
a Kerr medium, as described by Eqs.~\eqref{eq:HamiltonianKerr} and
\eqref{eq:QLEKerr}. This will be confirmed in the next section by
obtaining the quantum steady state of both systems numerically and
showing that the states of the optical mode are similar.

Many experimental realizations of cavity optomechanics are in the
resolved-sideband limit and fall into this category
\footnote{Our treatment neglects other possible nonlinear effects that could preclude the observation of instability in the upper branch. In particular, photothermal forces has proven important in experiments with silica microcavities as reported in Ref.~\onlinecite{Arcizet2009PRA80}}: micromechanical microwave
resonators \cite{Regal2008NPh,Rocheleau2010Nat,Teufel2011Nat,Massel2011Nat},
coated micromechanical resonators \cite{Groblacher2009Nat}, photonic crystal
cavities \cite{Chan2011Nat}, microspheres \cite{Park2009NPh}, and microtoroids \cite{Schliesser2008NPh,Verhagen2012Nat}.

\subsubsection{Unresolved sideband and small mechanical damping (IIb)}
\label{sec:IIb}

The critical occupation $n_c$ can be approximated in the limit of
a small sideband parameter $\omega_m/\kappa$ and large enough mechanical
quality factor, such that $1>\omega_m/\kappa > 1/Q_m$, as
\begin{equation}
\chi n_c = \frac{1}{4}\left(y + \sqrt{y^2 + 
\frac{\kappa/\omega_m}{8 \,Q_m}}\right)\:.
\label{eq:criticalMF3}
\end{equation}
If the bare detuning $\Delta_0$ is negative and exceeds the threshold value for
possible bistability, $y>\tilde{y}$, we obtain that $n_c \simeq
n_{\Delta}$. The upper branch turns unstable as soon as the effective
detuning $\Delta$ is positive, and for large bare red detuning, $-\Delta_0
\gg \kappa$, the upper branch is only stable on a small segment close
to its beginning.

In this regime we find several experimental implementations of optomechanics: 
ultracold atoms \cite{Murch2008NPh,Schleier-Smith2011PRL107,Brooks2012Nat},
suspended membranes \cite{Thompson2008Nat}, and coated mechanical resonators
\cite{Arcizet2006Nat,Kleckner2011OE}.

A simple interpretation of the critical mean-field occupation $n_c$ in
Eqs.~\eqref{eq:criticalMF2} and \eqref{eq:criticalMF3} can be
provided by considering the total mechanical damping
$\gamma_{\text{tot}} = \gamma_m +\Gamma_{\text{opt}}$,
where $\Gamma_{\text{opt}}$ is the additional mechanical damping induced
by coupling to the optical degree of freedom.
In the weak-coupling limit of linearized optomechanics, i.e., $g,\gamma_m <
\kappa$, this contribution is given by
$\Gamma_{\text{opt}}=-2\,\text{Im}\,\Sigma(\omega_m)$ where $\Sigma(\omega)
= -i g^2 \left[\chi_c(\omega)-\chi_c^{*}(-\omega)\right]$ is the so-called
optomechanical self-energy and
$\chi_c(\omega)=\left[\kappa/2-i(\Delta+\omega)\right]^{-1}$ the
optical susceptibility \cite{Marquardt2007PRL99}. In this case, the
condition $\bar{n} = n_c$ coincides with $\gamma_{\text{tot}} = 0$ in
both limits $\omega_m \lessgtr \kappa$.

\subsubsection{Very small sideband parameter}

In the regime where the sideband parameter is so small that
$\omega_m/\kappa \ll 1/Q_m$, the situation is different. The upper
branch is unconditionally stable as long as the detuning $y$ is not
too large, $y < \kappa/(\sqrt{32}Q_m \omega_m)$. For larger values of
$y$, an unstable segment of the upper branch develops, from the second
turning point $n_+$ up to some value $n'$ of the mean-field cavity
occupation given by
\begin{equation*}
\chi n' = y \left(\frac{1}{2} + Q_m\frac{\omega_m}{\kappa} 
+ \sqrt{\left(Q_m \frac{\omega_m}{\kappa}\right)^2 - \frac{1}{32 y^2}}\right).
\end{equation*}

The dynamical timescales of the two modes are different in this limit.
The optical mode adiabatically
follows the mechanical motion and produces an effective mechanical
potential with two stable equilibrium positions. However, as we have
seen in the previous paragraph, this picture holds only if $Q_m$ is
not too large compared to $\kappa/\omega_m$.

In this parameter regime, early experiments with hertz-scale
mechanical resonance frequencies enabled the first observations of
optical bistability and the related hysteresis cycle both in the optical
\cite{Dorsel1983PRL51} and the microwave domain \cite{Gozzini1985JOS}.

In low-finesse cavities, the optical field can create several stable minima in the
mechanical potential, a phenomenon sometimes referred to as multistability
\cite{Meystre1985JOS2,Meystre2007}. It has recently been observed with a
torsion balance oscillator acting as the moving mirror
\cite{Mueller2008PRA77}. This effect should not be confused with
\textit{dynamical} multistability \cite{Marquardt2006PRL96}, where
mechanical limit-cycle orbits of stable amplitudes arise due to parametric
instability.

\section{Optical bistability in the quantum regime}
\label{sec:equivalence}

\begin{figure*}
\centering
\includegraphics[width=0.93\linewidth]{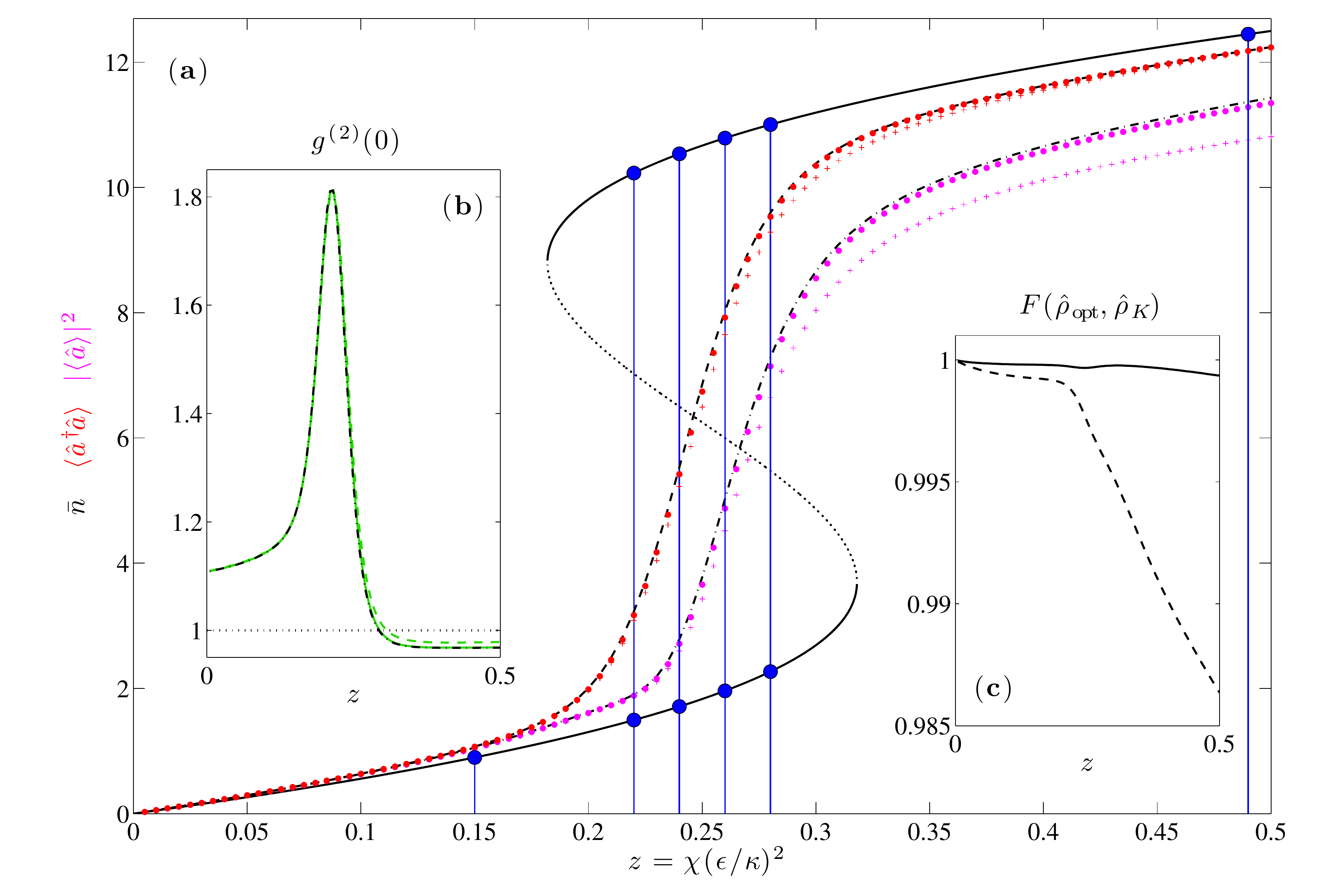}
\includegraphics[width=0.93\linewidth]{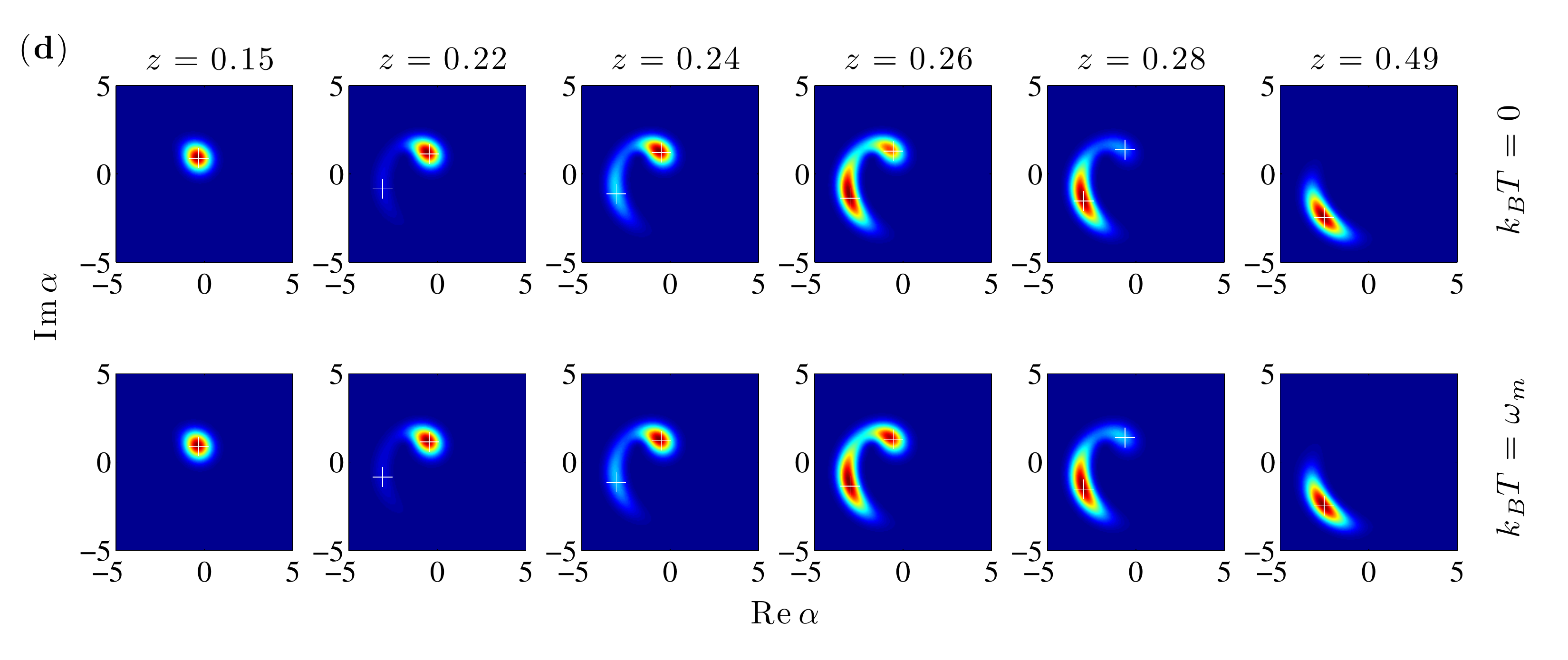}
\caption{(Color online) \textit{Optical bistability in the quantum
    regime}.  (a) Mean-field cavity occupation $\bar{n}$, with stable
  (black solid line) and unstable (black dotted line) branches,
  steady-state photon number $\langle \hat{a}^{\dagger} \hat{a}
  \rangle$ (red), and cavity amplitude $|\langle\hat{a}\rangle|^2$
  (purple) of the optomechanical system, as a function of the
  dimensionless driving power $z$.  The upper branch turns unstable
  outside the range of $z$ parameters we plot, beyond $z_c\simeq 92$
  and $n_c\simeq 42$. For comparison we also show $\langle
  \hat{a}^{\dagger} \hat{a} \rangle$ (black dashed line) and
  $|\langle\hat{a}\rangle|^2$ (black dash-dotted line) for the
  equivalent Kerr medium. For both systems, $y = -\Delta_0/\kappa =
  1.5$ and $\chi = 0.08$. The parameters of the optomechanical system
  are $\omega_m/\kappa=30$, $Q_m = 300$ (indicated by the black cross
  in Fig.~\ref{fig:fig2}), and $k_BT = 0$ (dots) or $k_BT=\omega_m$
  (crosses).  Inset (b) shows the second-order correlation function
  $g^{(2)}(0) = \langle \hat{a}^{\dagger} \hat{a}^{\dagger} \hat{a}
  \hat{a} \rangle / \langle \hat{a}^{\dagger} \hat{a} \rangle^2$ for
  the optomechanical system with $k_BT=0$ (green solid line) as well
  as $k_BT=\omega_m$ (green dashed line) and for the Kerr medium
  (black dash-dotted line).  The first and third curves are
  indistinguishable.  Inset (c) shows the overlap
  $F(\hat{\rho}_{\text{opt}},\hat{\rho}_K)$, as defined in
  Eq.~\eqref{eq:fidelity}, between the density matrices of the pure
  Kerr medium $\hat{\rho}_K$ and the reduced density matrix of the
  optomechanical system $\hat{\rho}_{\text{opt}}$, obtained by tracing
  out the mechanical degree of freedom from $\hat{\rho}$. The
  temperatures chosen are $k_BT=0$ (solid line) and $k_BT=\omega_m$
  (dashed line). (d) Wigner function $W_{\text{opt}}(\alpha)$ of the
  optical mode of the optomechanical system for six different driving
  powers $z$ and two different temperatures.  The white crosses
  indicate the mean-field amplitudes $\bar{a}$ of the stable branches.
  The values of $z$ are indicated by blue dots and lines in (a).}
\label{fig:fig3}
\end{figure*}

So far we have focused on the semiclassical regime, considering the
mean-field solutions as well as the effect of fluctuations around
them, and have identified the regime of parameters where the
optomechanical system and the Kerr medium exhibit similar behavior. In
the remainder, we want to confirm that the conclusions of this
approach also hold in the quantum limit. To this end, we compare the
quantum steady states of the optomechanical system and the Kerr
medium, obtained from numerical solutions of the quantum master
equations.

\subsection{Quantum master equations description of dissipation}

An alternative description of either the optomechanical system or the
Kerr medium can be given in the form of quantum master equations,
which describe the dynamics of their density operators $\hat{\rho}$,
respectively $\hat{\rho}_K$. This treatment is equivalent to the
quantum Langevin description given by Eqs.~\eqref{eq:QLE} and
\eqref{eq:QLEKerr}. Instead of using input noise operators $\hat{\xi}$
or $\hat{\eta}$, dissipation is taken into account with 
Lindblad dissipative terms.

The quantum master equation for the optomechanical system reads
\begin{equation}
\begin{aligned}
\frac{d \hat{\rho}}{dt} = \mathcal{L}\left[\hat{\rho}\right] 
= -i & \left[\hat{H}_0 + \hat{H}_d,\hat{\rho} \right]
+ \kappa\, \mathcal{D}_{\hat{a}}\left[ \hat{\rho} \right]
\\ +& (n_{\text{th}}+1)\gamma_m \, \mathcal{D}_{\hat{b}} \left[ \hat{\rho} \right]
+ n_{\text{th}} \gamma_m \, \mathcal{D}_{\hat{b}^{\dagger}} \left[ \hat{\rho} \right],
\label{eq:QME}
\end{aligned}
\end{equation}
where the dissipative terms have the standard form,
$\mathcal{D}_{\hat{o}}[\hat{\rho}] =
\hat{o}\,\hat{\rho}\,\hat{o}^{\dagger} - \frac{1}{2}
\left(\hat{o}^{\dagger}\hat{o}\,\hat{\rho} +
  \hat{\rho}\,\hat{o}^{\dagger}\hat{o}\right)$.

In the same way, the quantum master equation for the equivalent
Kerr medium is given by
\begin{equation}
\frac{d \hat{\rho}_K}{dt} = \mathcal{L}_K\left[\hat{\rho}_K\right] 
= -i \left[ \hat{H}_K + \hat{H}_d,\hat{\rho}_K \right] 
+ \kappa \mathcal{D}_{\hat{a}} \left[ \hat{\rho}_K\right].
\label{eq:QMEKerr}
\end{equation}

The steady-state density operators are found from the
numerical solutions of $\mathcal{L}[\hat{\rho}] = 0$ and
$\mathcal{L}_K[\hat{\rho}_K] = 0$, respectively.

\subsection{Comparison of the quantum steady states}

To corroborate the fact that the optomechanical system behaves like an
effective Kerr medium, we compare the quantum steady states of both
systems for parameters that lead to bistable behavior. To this end,
we calculate the photon number $\langle \hat{a}^{\dagger} \hat{a}
\rangle$, the cavity amplitude $|\langle\hat{a}\rangle|^2$, and the
second-order correlation function
\begin{equation*}
g^{(2)}(0) = \frac{\langle \hat{a}^{\dagger} \hat{a}^{\dagger} \hat{a}\hat{a} \rangle}
{\langle \hat{a}^{\dagger} \hat{a} \rangle^2}\:,
\end{equation*}
which describes fluctuations in the photon number. We also
characterize the similarity between the optomechanical system and the
Kerr medium with the help of the overlap
\begin{equation}
F \left(\hat{\rho}_{\text{opt}},\hat{\rho}_K \right) 
= \text{Tr} \left[\sqrt{\sqrt{\hat{\rho}_K} \hat{\rho}_{\text{opt}} 
\sqrt{\hat{\rho}_K} } \right] \:,
\label{eq:fidelity}
\end{equation}
where $\hat{\rho}_{\text{opt}}$ is the reduced density matrix of the
system, obtained by tracing out the mechanical degree
of freedom from $\hat{\rho}$. Finally, we investigate the Wigner
distribution function of the optical mode, which reads
\begin{equation*}
W_{\text{opt}} (\alpha) = \frac{1}{\pi^2} \int d^2\lambda\, 
\text{Tr}\left[ \hat{\rho}_{\text{opt}}\,
  e^{\lambda(\hat{a}^{\dagger}- \alpha^{*})
- \lambda^{*}(\hat{a}-\alpha)} \right]\:.
\end{equation*}

The steady states of both systems are compared for a constant detuning
above the bistability threshold, $y > \tilde{y}$, and  as a function of the
driving power $z$. In this configuration the mean-field cavity occupation
$\bar{n}$ forms a characteristic $S$-shaped curve.

The results are presented in Fig.~\ref{fig:fig3}. In the upper panel,
we show the mean-field cavity occupation $\bar{n}$, the photon number
$\langle \hat{a}^{\dagger}\hat{a}\rangle$, and the cavity amplitude
$|\langle\hat{a}\rangle|^2$ for both the optomechanical system, with
zero and finite temperature of the mechanical bath, as well as for the
equivalent Kerr medium. The two insets show the second-order
correlation $g^{(2)}(0)$ and the overlap $F
\left(\hat{\rho}_{\text{opt}},\hat{\rho}_K \right)$. The lower panel
of Fig.~\ref{fig:fig3} shows the optical Wigner density function of
the optomechanical system.

At low driving power before entering the region of bistability,
$z<z_-$, the state of the optical mode is rather well described by a
coherent state in both systems, as $\langle \hat{a}^{\dagger}
\hat{a}\rangle = |\langle \hat{a} \rangle|^2 \simeq \bar{n}$.

In the range of driving power where two stable mean-field solutions
exist, $z_-<z<z_+$, the master equations~\eqref{eq:QME} and
\eqref{eq:QMEKerr} have \textit{unique} quantum steady states.  Thus,
instead of any bistable behavior, a transition of $\langle
\hat{a}^{\dagger}\hat{a}\rangle$ and $|\langle\hat{a}\rangle|^2$, from
the lower to the upper branch, occurs, as the driving power $z$
increases.  Simultaneously, both systems show large fluctuations in
the photon number, $g^{(2)}(0) > 1$. Such behavior, in the regime
where the MFEs lead to bistability, is well-known from the Kerr medium
\cite{DrummondWalls1980JPA13}.

In this regime, the Wigner function $W_{\text{opt}}(\alpha)$, shown in
the lower part of Fig.~\ref{fig:fig3}, exhibits two separate lobes
peaked at the mean-field amplitudes, $\alpha \simeq \bar{a}$.  This is
another well-known feature of the Kerr medium \cite{Risken1987PRA35,
*Vogel1989PRA39,Vogel1990PRA42} and 
shows how classical bistability persists in the quantum regime.
The two lobes are distinguishable if the phase-space
separation of the two stable mean-field amplitudes $\bar{a}$ is larger
than the fluctuations around them, which is satisfied here 
since $\chi \ll 1$.
Since $W_{\text{opt}} > 0$ everywhere, the optical mode can be
regarded as an incoherent statistical mixture of two states with
different amplitudes and non-Gaussian fluctuations. As the driving
power $z$ increases from $z_-$ to $z_+$, the relative weights of the
lobes continuously change from the lower branch to the upper one,
describing the shift in probability for the system to be found in one
or the other.  This effect is robust to finite
temperature of the mechanical environment.

The particular situation where the two stable branches are
approximately equally likely ($z\simeq 0.26$ for $k_B T = \omega_m$)
would enable the observation of noise-induced switching between
both branches \cite{Rigo1997PRA55,Kerckhoff2011OE19} and constitute a
clear signature of the nonlinear interaction between the optical and
mechanical mode.

At higher driving power, $z>z_+$, when the MFEs have only one
solution, both the optomechanical system and the Kerr medium exhibit
sub-Poissionian statistics, $g^{(2)}(0) < 1$. Photon blockade in
optomechanical systems has already been predicted
for $\chi >1$ \cite{Rabl2011PRL107}. In our case, photon blockade is not very
pronounced: we chose $\chi \ll 1$ to have bistable mean-field
solutions that are appreciably distant in phase space. For the
parameters of Fig.~\ref{fig:fig3}, this effect is slightly suppressed
even further due to the finite-temperature bath, $n_{\text{th}}>0$.

At various points of the paper, we have already demonstrated that the
optomechanical system can be regarded as an effective Kerr medium in
some range of parameters that we have specified. In particular, in the
present section we have shown numerically that both systems exhibit
the same features. For example, the photon number and the second-order photon
correlation function follow the same parameter dependence, the Wigner
function has a two-lobe structure, and both systems show photon
blockade.  As a further strong confirmation of this equivalence, we compare
the states $\hat{\rho}_{\text{opt}}$ and $\hat{\rho}_K$ of
the optical field in both systems.  As can be seen in inset (c) of
Fig.~\ref{fig:fig3}, their overlap $F$ is close to 1 even at a finite
thermal occupation of the mechanical mode. All of these calculations
clearly establish the equivalence of the optomechanical system and a
Kerr medium in the appropriate parameter range.

\section{Conclusion}

The mean-field equations for the optical mode of a dispersively
coupled optomechanical system agree with those of a Kerr medium, a
paradigmatic quantum optics system whose nonlinearity induces optical
bistability.  This raises the question of whether and under which
conditions the two systems can be considered to be equivalent.  We
have therefore compared the optical bistability in an optomechanical
system and a Kerr medium.  A stability analysis of the mean-field
solutions reveals differences between the two systems: the upper
branch of an optomechanical system can become unstable due to position
fluctuations of the mechanical degree of freedom. We have identified
the regime of parameters where the two systems are equivalent.
Corroborating this semiclassical approach, we have shown that the
(optical) quantum steady states of both systems, obtained numerically,
show large overlap. Our results clarify when an optomechanical system
can be used as a Kerr nonlinearity in applications of quantum optics
and quantum information.

\section*{Acknowledgements}
We would like to thank D.~Vitali for interesting discussions.
This work was financially supported by the Swiss NSF, the NCCR
Nanoscience, and the NCCR Quantum Science and Technology.

\end{document}